 \documentclass[final,onefignum,onetabnum]{siamart171218}

\usepackage{lipsum}
\usepackage{amsfonts}
\usepackage{graphicx}
\usepackage{epstopdf}
\ifpdf
  \DeclareGraphicsExtensions{.eps,.pdf,.png,.jpg}
\else
  \DeclareGraphicsExtensions{.eps}
\fi

\newsiamremark{remark}{Remark}
\newsiamremark{hypothesis}{Hypothesis}
\crefname{hypothesis}{Hypothesis}{Hypotheses}
\newsiamthm{claim}{Claim}

\headers{
Large scale simulation of fracture propagation
}{
P. Zulian, 
A. Kopani\v c\'akov\'a, et al.
}

\title{Large scale simulation of pressure induced phase-field fracture propagation using Utopia
\thanks{Submitted to the editors DATE.
\funding{
P.Z., M.G.C.N. and R.K. thank the project ``Forecasting and Assessing Seismicity and Thermal Evolution in geothermal Reservoirs'' (FASTER) founded by Platform for Advanced and Scientific Computing (PASC). Additionally, this research is part of the activities of the Swiss Centre for Competence in Energy Research on Supply of Electricity (SCCER-SoE) and the Future Swiss Electrical Infrastructure (SCCER-FURIES), which is financially supported by the Swiss Innovation Agency (Innosuisse - SCCER program).
A.K. and R.K. thank the project "Large-scale simulation of pneumatic and hydraulic fracture with a phase-field approach" (No.:154090), founded by the Swiss National Science Foundation (SNF)
and the project  "Reliable Simulation Techniques in Solid Mechanics. Development of Non-standard Discretization Methods, Mechanical and Mathematical Analysis" (No.:SPP1748) founded by the Deutsche Forschungsgemeinschaf (DFG).
}}}

\author{
Patrick Zulian \footnotemark[5]
\thanks{Authors contributed equally}
\and 
Alena Kopani\v c\'akov\'a  \footnotemark[5]
\footnotemark[2]
\and 
Maria Giuseppina Chiara Nestola \footnotemark[5]  \thanks{Institute of Geochemics and Petrology, ETH Zurich, Switzerland
  (\email{maria.nestola@erdw.ethz.ch})}
\and 
Andreas Fink \thanks{Swiss National Supercomputing Centre (CSCS), ETH Zurich, Switzerland
  (\email{andreas.fink@cscs.ch}, \email{nur.fadel@cscs.ch}, \email{joost.vandevondele@cscs.ch} ).}
\and 
Nur Aiman Fadel\footnotemark[4]
\and 
Joost VandeVondele\footnotemark[4]
\and 
Rolf Krause
\thanks{Institute of Computational Science, Universit\`a della Svizzera italiana, Lugano, Switzerland
  (\email{patrick.zulian@usi.ch}, \email{alena.kopanicakova@usi.ch}, \email{nestom@usi.ch}, \email{rolf.krause@usi.ch} ).}}

\usepackage{amsopn}

\makeatletter
\newcommand*{\addFileDependency}[1]{
  \typeout{(#1)}
  \@addtofilelist{#1}
  \IfFileExists{#1}{}{\typeout{No file #1.}}
}
\makeatother

 \usepackage{bm}
\usepackage{calrsfs}
\DeclareMathAlphabet{\pazocal}{OMS}{zplm}{m}{n}
\usepackage{tikz}
\usetikzlibrary{arrows}
\usepackage{algorithm}
\usepackage{algpseudocode}
\usepackage{pgfplots}
\usepackage{pgfplotstable}

\algnewcommand\algorithmiconput{\textbf{Constants:}}
\algnewcommand\algorithmicinput{\textbf{Input:}}
\algnewcommand\algorithmicoutput{\textbf{Output:}}
\algnewcommand{\algorithmicgoto}{\textbf{go to}}%

\algnewcommand\Constants{\item[\algorithmiconput]}
\algnewcommand\Input{\item[\algorithmicinput]}%
\algnewcommand\Output{\item[\algorithmicoutput]}%
\algnewcommand{\Goto}[1]{\algorithmicgoto~\ref{#1}}%

\def \pi{{\mathbf{p}_{i}}}          
\def \0{\mathbf{0}}         %

\def \x{{\mathbf{x}}}           
\def \s{{\mathbf{s}}}           
\def \t{\mathbf{t}}                        
\def \n{\mathbf{n}}                        
\def \R{{\mathbb{R}}}           
\def \N {\mathbb{N}}            
\def \g{{\mathbf{g}}}           %
\def \H{{\mathbf{H}}}           %

\def \l{{\mathbf{l}}}           
\def \u{{\mathbf{u}}}           
\def \eps{\bm{\varepsilon}}     

\def \Rev{{\mathbf{R}}}         
\def \P{{\mathbf{P}}}           
\def \I{{\mathbf{I}}}           

\definecolor{myblack}{RGB}{53, 53, 53}
\definecolor{myblue}{RGB}{40, 75, 99}
\definecolor{myred}{RGB}{192, 50, 33}
\definecolor{myyellow}{RGB}{255, 166, 48}
\definecolor{mywhite}{RGB}{240, 237, 238}
\definecolor{mygreen}{RGB}{0, 102, 0}

\definecolor{green1}{RGB}{9, 82, 86}
\definecolor{green2}{RGB}{8, 127, 140}
\definecolor{green3}{RGB}{6, 167, 125}
\definecolor{green4}{RGB}{79, 109, 122}
\definecolor{green5}{RGB}{192, 214, 223}
\definecolor{violet}{RGB}{26,69,131}

\ifpdf
\hypersetup{
  pdftitle={Large scale simulation of pressure induced phase-field fracture propagation using Utopia},
  pdfauthor={Patrick Zulian,
Alena Kopani\v c\'akov\'a,
Maria Giuseppina Chiara Nestola,
Andreas Fink,
Nur Aiman Fadel,
Joost VandeVondele,
and
Rolf Krause}
}
\fi

\begin{document}

\maketitle

\begin{abstract}
Non-linear phase field models are increasingly used for the simulation of fracture propagation models. The numerical simulation of fracture networks of realistic size requires the efficient parallel solution of large coupled non-linear systems. Although in principle efficient iterative multi-level methods for these types of problems are available, they are not widely used in practice due to the complexity of their parallel implementation.

Here, we present Utopia, which is an open-source C++ library for parallel non-linear multilevel solution strategies. Utopia provides the advantages of high-level programming interfaces while at the same time a framework to access low-level data-structures without breaking code encapsulation.
Complex numerical procedures can be expressed with few lines of code, and evaluated by different implementations, libraries, or computing hardware.
In this paper, we investigate the parallel performance of our implementation of the recursive multilevel trust-region (RMTR) method based on the Utopia library.
RMTR is a globally convergent multilevel solution strategy designed to solve non-convex constrained minimization problems.
In particular, we solve pressure-induced phase-field fracture propagation in large and complex fracture networks.
Solving such problems is deemed challenging even for a few fractures, however, here we are considering networks of realistic size with up to 1000 fractures.
\end{abstract}




\section{Introduction}
Fractures and fracture networks do strongly affect the hydraulic and mechanical response of the underground.  This is of particular relevance for geothermal technologies, 
which aim at producing electricity from deep geothermal resources by enhancing the
permeability of a geothermal reservoir to obtain a sufficiently large heat flux on interior
surfaces~\cite{chen2018evaluation,samin2019hybrid,sokolova2019multiscale}. 
This is usually done by hydraulic stimulation. Unfortunately, the  Enhanced Geothermal Systems (EGSs) resulting from this stimulation process might face several challenges such as induced seismicity events, possibly inhibiting industrial development. The development of dedicated simulation tools allows for substantial savings in the cost of experiments needed to improve the design of hydraulic stimulations and their overall performance~\cite{collignon2020evaluating}.

In numerical simulations, realistic fracture networks are usually challenging 
to represent with a discrete geometry (i.e., a mesh), or even impossible at the macro-scale.
Moreover, the use of fine meshes, necessary to produce accurate results,  makes the simulations numerically and computationally demanding, since they give rise to large scale simulation in terms of both, domain size and resolution (number of degrees of freedom).
Hence, fractures and fracture networks lead to particular modeling challenges, such as:
\begin{enumerate}
\item integrating heterogeneities, like natural fractures, in large computational domains
(typically of the order of kilometers) with high resolution;
\item  representing and modeling fracture initiation and propagation.
\end{enumerate}

The broad family of numerical models for fractured porous media can be classified into two  different approaches: (1) the discrete-fracture-networks model (DFN)~\cite{hyman2015} where the matrix is neglected, and (2) the discrete-fracture-matrix model (DFM) where the matrix and its coupling with the fracture network is explicitly represented. Here, due to their huge length-to-width ratio, fractures are usually represented as lower-dimensional manifolds embedded in the porous matrix.
Two different numerical approaches have been adopted for DFMs, mainly classified as conforming and non conforming. 
The first class requires computational meshes where fracture and matrix share the same nodes on the interface ~\cite{flemisch_2011}. 
The fully conforming approach moves all complexity to the meshing procedure which can be very challenging and time consuming~\cite{cacace2015}. 
The second class of numerical schemes removes the requirement of mesh conformity and makes use of 
specific techniques such as mortar methods~\cite{schaedle20193d,fumagalli2019conforming,chernyshenko2020unfitted}, and extended finite element discretizations~\cite{vahab2019x} to ensure the continuity of pressure and fluxes in intersecting fractures.

Fractures with considerably large aperture width may not be well represented by lower-dimensional models. Hence, equi-dimensional approaches are required to geometrically account for the fracture aperture. In this regard, in~\cite{formaggia2018analysis} a Mimetic Finite Difference (MDF) Method has been introduced to preserve the quality of the solution for highly distorted grids, whereas in ~\cite{berre2020verification} an adaptive mesh refinement has been adopted to increase the accuracy of the solution in the fracture zone. 

Another important ingredient concerns the numerical modelling of  seismic events. Adaptive 
hierarchical fracture approaches, as well as upscaling and multiscale methods for EGS \cite{karvounis2016adaptive,kiraly2016validating,chung2018non,praditia2018multiscale,berrone2019parallel} 
have been proposed for modeling the interaction between the natural existing fracture network and the pressure-induced hydraulic fractures. Here, fractures are usually represented as sharp interfaces that are added to the pre-existing network if the stress 
computed on the corresponding hypocenters satisfies a given failure criterion. However, 
the computation of the stress relies on model rules which are not physics-based but rely on the use of stochastic approaches deduced by experimental results.

\subsection{Phase-field fracture approaches}
Phase-field approaches for fracture allow for modelling and  simulating the fracture initiation, propagation, and interaction without the need of explicitly representing the fracture surface.
 The basic idea of this method is to model systems with sharp interfaces or fractures using a continuous variable, called the phase-field, that allows for incorporating the presence of fractures into a given system through a \textit{smooth} transition between two states, i.e. damaged and not damaged
 Moreover, combining phase-field approaches with elastodynamical or poroelastic models allows for computing both, the evolution of cracks and the stress distribution around them.

The first numerical implementation of a variational phase-field approach was presented in~\cite{Bourdin2007}.
Miehe et al.~\cite{miehe2010thermodynamically, miehe2010phase} enhanced the underlying mathematical model and introduced thermodynamically consistent, rate-independent formulation.
Since then, the phase-field approach has become popular in the literature and has been extended in many directions, including
dynamic models \cite{bourdin2011time,schluter2014phase},
shells and plates approaches \cite{amiri2014phase, ambati2016phase},
generalization to large deformations \cite{delpiero2007variational,hesch2014thermodynamically,bilgen2019detailed},
adaptive fourth-order models \cite{goswami2020adaptive, weinberg2017high},
hybrid schemes \cite{goswami2020adaptive},
or anisotropic models for a fracture of fiber-reinforced matrix composites \cite{denli2020phase}.
Application of phase-field approaches to fracture initiation and propagation include also cohesive fractures \cite{vignollet2014phase, conti2016phase} and hydraulic fracturing \cite{bilgen2017phase,mikelic2015quasi,Heider2016}.
For further details, we refer the interested reader to the review provided in~\cite{delorenzis2020numerical}.

\subsubsection{Towards large scale simulations}
Although phase-field models and their coupling with elastodynamics thermodynamics  are very popular in several multiphysics simulations, spanning from particle growth~\cite{moelans2006phase} to ductile fracture~\cite{ambati2015review} and geoscience~\cite{mikelic2015quasi} its applicability is currently  limited to  small scale problems due to its burdensome computational cost.
Firstly, high-resolution meshes are required to resolve the localized damage, which leads to large scale simulations with a huge number of degrees of freedom.
Secondly, solving the resulting problems numerically is challenging as it requires the minimization of a  non-convex energy functional.
As a consequence, standard solution strategies, such as Newton's method, often fail to converge.
The alternate minimization~\cite{bourdin2011time} has been widely adopted to solve phase-field fracture problems~\cite{farrell2017linear, li2016gradient, amor2009regularized, wu2017phase, wu2018length}. The main idea behind this method is to minimize the energy functional successively for displacements and phase field. This gives rise to two convex minimization subproblems, which are then alternatingly solved until convergence is reached. Although solving the convex sub-problems is fairly straightforward, the overall convergence speed of
the method can be erratic~\cite{farrell2017linear}.
Moreover, the scalability properties of this approach are also limited, as the number of variables, and consequently, the size, of the two sub-problems differs.

In this regard,  monolithic approaches can be computationally more efficient than an alternate minimization and several attempts have been made to enhance their robustness, which includes path-following strategies~\cite{singh2016fracture}, line-search methods~\cite{GERASIMOV2016276}, primal-dual algorithms~\cite{heister2015primal}, modified Newton's method~\cite{wick2017modified}, quasi-Newton's method based on BFGS~\cite{wu2020b}, or Fast Fourier Transform (FFT) solution strategies~\cite{chen2019fft}.

However, the applicability of these approaches to large scale problems is mainly limited by the use of direct linear solvers for 
the solution of the arising  linear systems.
To this aim, multilevel strategies have been employed as an inner linear solver, due to their optimal complexity.  In particular, a geometric multigrid method was applied in \cite{bilgen2017phase} showing scalability up to $300$ processes, while matrix-free multigrid was used in \cite{jodlbauer2019matrix}, demonstrating scalability up to $128$ cores.
Alternative approaches, based on truncated non-smooth non-linear monotone multigrid, were used in~\cite{kienle2018efficient}, where authors obtained a significant improvement in terms of computational time, but the parallel performance was not reported.
More recently,for phase field models a variant of the
nonlinear multilevel method based on the trust region method, called Recursive Multilevel
Trust region (RMTR)~\cite{Gratton2008recursive, gratton2008_inf, Gross2009}, has been developed in~\cite{kopanicakova20a}. RMTR for phase field ensures global  convergence and has been shown to scale up to $300$ processes.

Even though large scale phase-field fracture approaches can be found in the literature, very few attempts have been made to apply them to large scale problems inspired by real-world applications, see for example \cite{wick2016fluid, yoshioka2016variational, mollaali2019numerical}.
In this work, we developed a phase-field fracture simulation code in order to model the fracture propagation and interaction processes of large scale fracture networks.
In particular, motivated by the promising properties of the RMTR method, shown in~\cite{kopanicakova20a}, we extended the approach to complex scenarios with hundreds of fractures in three-dimensions and with thousands of fractures in two-dimensions. To our knowledge, this is the first time that the phase-field approach 
is employed for such complex, large scale scenarios. 

\subsubsection{Source codes}
Compared to discrete-fracture approaches, the finite element implementation of phase field models is relatively straightforward.
Most of the results reported in the literature rely heavily on in-house academic codes, based for example on environment Matlab~\cite{nguyen2015phase, nguyen2017large, hesch2017framework}.
First commercial implementations appeared in software such as Abaqus~\cite{liu2016abaqus, molnar20172d, msekh2015abaqus}  and COMSOL~\cite{zhou2018phase}.
More recently, several open source codes were reported in the literature, for example~\cite{farrell2017linear} and~\cite{li2016gradient}  use the finite element framework FEniCS~\cite{logg2007} in order to implement a quasi static and dynamic model for brittle fractures, respectively.
The implementation documented in~\cite{heister2015primal, klinsmann2015assessment} relies on the package Deal II.~\cite{BangerthHartmannKanschat2007} and supports adaptive mesh refinement strategies.
 The MOOSE environment~\cite{gaston2009moose} served as a base for the implementation reported in~\cite{chakraborty2016phase, chakraborty2016multi}.
The results obtained in~\cite{kuhn2015degradation, steinke2016comparative} were produced using FEAP~\cite{feap}.
Additionally, the JIVE framework~\cite{jive}  was utilized in~\cite{may2015numerical}, while the package NUTIL~\cite{van2018nutils} was used in~\cite{singh2016fracture} instead.
A GPU implementation was presented in~\cite{ziaei2016massive}, where the authors demonstrate a speedup factor of $12$ for simulations with around $2.5$ million degrees of freedom (dofs).
A thread scalable implementation based on the Kokkos library~\cite{edwards2014} was presented in~\cite{tupek2016cohesive} for cohesive fracture.


\subsection{The Utopia library} \label{sec:utopialib}
Due to power and thermal restrictions, CPUs have almost reached their limits when it comes to high-performance data processing. This gave rise to new technologies like GPGPU (general purpose graphics processing units) and Tensor Cores, but also lead to the enhancement of older technologies such as  FPGAs.
%
With such new developments, scientific-computing software libraries need to be constantly updated or rewritten. For instance, the advent of GPGPU induced new programming paradigms and new languages such as Cuda~\cite{nickolls2008} and
OpenCL~\cite{opencl08}, which led to the creation of new software libraries such as CuBLAS~\cite{cublas2008} and ViennaCL~\cite{rupp2016}.
Keeping up with such new technologies may cause small to significant changes in the code of
software-applications such as non-linear solution strategies, finite element analysis, and data-analysis.
However, the related high-level algorithms implemented in the application code should not have to change.

For this reason, one solution is to develop application code  on top of a portable interface that fits many current and possibly future requirements (e.g., PETSc~\cite{petsc-efficient,petsc-user-ref},  Trilinos{}~\cite{heroux2003}, and Kokkos~\cite{edwards2014}).
For instance, software libraries such as Deal.II~\cite{BangerthHartmannKanschat2007}, LibMesh~\cite{kirk2006}, Dune~\cite{dune2016}, and MOOSE~\cite{gaston2009moose} rely on high level abstractions on top of existing linear algebra and non-linear solution strategies codes, and allow choosing, to some degree, the underlying implementation.
However, a clear separation of frontend programming and the backend implementation would help in keeping up with even new technologies or upcoming and yet unknown paradigm shifts.
A best-case scenario allows us to never touch the frontend code and implement new backends based on these new technological advancements or even mix multiple backends and exploit the best parts of each world.

To this end, a possible solution is to exploit scripting facilities for completely decoupling the application behavior from its actual implementation. This solution has the advantage of hiding the complexity of parallel software to which the average, casual or opportunistic~\cite{brandt2008}, user is not supposed to be exposed.
The idea is that the scripting code is translated to behavior which is implemented in another lower-level language.
This enables users to write a few lines of very powerful code without the overhead of learning how to use new complex parallel scientific codes.
A very specific form of scripting language is usually referred to as \emph{domain specific language} (DSL). This specificity, while reaching the aforementioned objectives, has a twofold advantage. First, it enables a simple description of a specific problem since most implementation details can be hidden.
Second, it allows exploiting complex functionalities and performance critical optimizations.
Notable examples related to finite element software, are Fenics' unified form language~\cite{logg2007,rathgeber2016}, FreeFEM~\cite{hecht2012}, and Liszt~\cite{devito2011}.

In DSLs lower-level abstractions are purposefully inaccessible because the actual algorithms are implemented in a different language, such as C++.
This is a problem when a DSL misses a functionality, since adding it would require accessing the underlying back-end which may be either closed source or very complex. In contrast, embedded domain-specific languages (eDSL) (e.g., CULA~\cite{humphrey2010}, Feel++~\cite{prudhomme2020}, OpenFOAM~\cite{weller1998}, Sundance{}~\cite{long2010}) use the same language and compiler for both the ``scripting'' layer and the implementation of the back-end. For this reason, eDSLs have the opportunity to provide the right balance between abstraction and direct access to the back-end data-types and algorithms.

In this paper, we use the open-source C++ library Utopia \cite{utopiagit}, which currently
provides a uniform interface to the PETSc algebra, and Tpetra from the Trilinos library.
The first goal of Utopia is the separation of model and computation (similar to DSLs) and its main purpose is advanced parallel algebra (linear and non-linear).
By exploiting meta-programming facilities in combination with expression templates~\cite{iglberger2012,veldhuizen1995}, Utopia can easily be integrated with any other existing parallel algebra library, hence it is mostly independent from technological changes.

The second goal is to provide a uniform interface to lower-level technologies (e.g., Kokkos, RAJA~\cite{beckingsale19performance}, or SyCL~\cite{bader2019sycl}).
In fact, the Utopia library is designed and developed for providing a balance between abstraction and low-level access without sacrificing performance. It aims at an organic integration with existing code without creating barriers between abstractions and implementation. High level and lower level abstractions, as well as raw data are accessible to the user at any time. This allows users to extend their code with possibly missing functionalities by manipulating lower-level abstractions and eventually even the low-level data (and back-end) directly.
The flexible design of Utopia allows for adding these features in a straightforward way to future releases without changing the high-level interfaces.

The third goal is to reduce the overhead of the front-end and allow to exploit available functionalities of the different back-ends as good as possible.
To this end, Utopia exploits static polymorphism, so that no performance-overhead associated with virtual tables is introduced, and specific evaluation routines can be specialized by exploiting partial/full specialization.

Our implementation of the finite element models for phase-field targets CPU architectures, hence we use the Utopia/PETSc based back-end in combination with the PETSc DM package for our computations.


\subsection{Contributions and content of the paper} \label{sec:contributions}
The five main contributions of this article are:
\begin{enumerate}
    \item the first introduction of the  open-source C++ library Utopia \cite{utopiagit};
    \item efficient open-source finite element code for phase-field fracture simulations;
    \item the only parallel open-source code of the RMTR method, an efficient globally convergent nonlinear multilevel solution strategy for non-convex constrained minimization problems;
    \item large scale simulations of pressure-induced fracture propagation of stochastic fracture networks, considering realistic and complex scenarios up to $1000$ fractures;
    \item strong and weak scaling studies up to $12\,288$ MPI processes and $1.05 \times 10^9$ degrees-of-freedom of the proposed algorithmic framework and its CPU-tailored implementation using Utopia;
\end{enumerate}

We start by describing the pressure-induced phase-field fracture model (\cref{sec:pf_frac}), and the recursive multilevel trust-region strategy (\cref{sec:rmtr}), adopted to solve the arising nonlinear systems.
Next, we provide an overview of our software and a detailed description of the developed code (\cref{sec:implementation}).
Then, we validate the implementation of the phase-field fracture model and present numerical experiments with complex fracture networks for applications in geoscience (\cref{sec:experiments}).
Furthermore, we demonstrate the strong and weak scaling performance properties of our code using Piz Daint super-computing machine (\cref{sec:perf}).
Finally, we provide concluding remarks and describe future plans (\cref{sec:conclusion}).


\section{Pressure induced phase-field fracture model}
\label{sec:pf_frac}
In this section, we briefly review pressure induced fracture processes modeled using the second-order phase-field formulation for brittle fracture.
Our presentation focuses on the  quasi-static time-discrete setting.
A pseudo-time step $t=1, \dots, T$, is used to index the deformation state in the loading process.
We denote the computational domain by $\Omega \in \R^d, d=2,3$, representing a $d$-dimensional solid with internal fracture $C \subset \R^{d-1}$, which evolves during the loading process.
The boundary $\partial \Omega$ of the domain $\Omega$ is decomposed into two non-overlapping parts, $\Gamma_D$, $\Gamma_N$, where Dirichlet and Neumann boundary conditions are prescribed, respectively.
Additionally, we set $\partial \Omega_N = \Gamma_N \cup \partial C$.

In this work, we assume that the body $\Omega$ shows linear elastic behaviour, with the strain energy density function defined as: 
$\psi_e( \eps(\u) ):= 0.5 \lambda (\text{tr}(\eps(\u)))^2 + \mu \eps(\u):\eps(\u)$, where $\mu, \lambda$ are the Lam\`e parameters,
$\u: \Omega \rightarrow \R^d $ represents the displacement vector field and $\eps(\u) := \text{sym}(\nabla \u)$ is the strain tensor.
Furthermore, we prescribe a given pressure $p: \Omega \rightarrow \R$, over the domain $\Omega$ to only induce fracture propagation.
Here we remark, that this work focuses only on the fracture propagation, i.e. we assume that pressure $p$ is given a priori.
We remark that in order to improve the reliability of the phase-field fracture model, it could be coupled with  poroelasticity equations such as the Biot's equations \cite{mikeli2015}.
This would allow for simulating induced hydraulic fracturing in a poro-elastic medium rather than in an elastic medium.

\subsection{Variational approach to fracture}
The variational approach proposed by Francfort and Marigo~\cite{francfort1998revisiting} formulates brittle fracture as a minimization problem for of an energy functional consisting of the elastic energy of the cracked solid, the energy dissipated in the fracture, and the traction forces; thus
\begin{equation}
\begin{aligned}
E(\u, C, p) := &\int_{\Omega \setminus C} \psi_e( \eps(\u) ) \ d\Omega +  \pazocal{G}_c S^{d-1}(C)
- \int_{\partial_N \Omega} \bar{\t} \cdot \u \ ds,
\label{eq:energy_not_reg}
\end{aligned}
\end{equation}
where $\pazocal{G}_c > 0$ denotes fracture toughness and $\bar{\t}$ stands for the traction forces.
The symbol $S^{d-1}(C)$ in \cref{eq:energy_not_reg} denotes the Hausdorff surface measure of fracture set $C$, i.e. $S^{d-1}(C)$ represent length or the surface area of fracture $C$, when $d=2,3$, respectively.
Note, that the traction forces $\bar{\t}$ constitute of two parts
\begin{align}
\int_{\partial_N \Omega} \bar{\t} \cdot \u \ ds = \int_{\Gamma_N} \bar{\t}_{\Omega} \cdot \u \ ds -  \int_{\partial C} p \ \n \cdot \u \ ds,
\label{eq:traction_forces}
\end{align}
where $ \bar{\t}_{\Omega}$ is traction force applied at the domain boundary $\Gamma_N$ and $\n$ is unit vector normal to the fracture surface.
The last term in \cref{eq:traction_forces} represents a force introduced by the pressure $p$ inside of the fracture, which is applied on a surface.

The direct minimization of the energy functional~\cref{eq:energy_not_reg} is computationally prohibitive, as the fracture surface $C$ is not known a priori.
Even in the incremental settings, formulation~\cref{eq:energy_not_reg}  requires the precise tracking of a moving fracture surface, which leads to computationally demanding algorithms.
To overcome this difficulty, Bourdin et. al~\cite{Bourdin2007} propose to utilize a regularization strategy initially developed by Ambrosio and Torelli \cite{ambrosio1992approximation} for image-segmentation.
The regularization strategy introduces a smooth scalar field, called phase-field $c: \Omega \rightarrow  [0,1]$, which characterizes the material state of the domain $\Omega$.
In particular, the value $c=0$ indicates an intact solid, $c=1$ denotes the fractured or broken state, while $c \in (0,1)$ constitute smooth transition zones between the two limit states.
Using the phase-field variable $c$, we can replace the fracture energy in~\cref{eq:energy_not_reg} by its volumetric approximation, thus
\begin{align}
\pazocal{G}_c S^{d-1}(\Gamma) \approx \frac{\pazocal{G}_c}{c_w} \bigg( \frac{w(c)}{l_s} + l_s \left| \nabla c \right|^2  \bigg)  d\Omega,
\label{eq:pf_approx}
\end{align}
where the length-scale parameter $l_s$ controls the thickness of the transition zone between the material states.
The function $w$ defines a decaying profile of the phase-field $c$, while $c_w:=4 \int_{0}^1 \sqrt{w(c)} \ dc$ is an induced normalization constant.
Taking into account \cref{eq:pf_approx}, we can reformulate \cref{eq:energy_not_reg} as
\begin{align}
    E(\u, c, p) := &\int_{\Omega} g(c) \ \psi_e( \eps(\u)) +   \frac{\pazocal{G}_c}{c_w} \bigg( \frac{w(c)}{l_s} + l_s \left| \nabla c \right|^2  \bigg)  d\Omega
    -  \int_{\partial_N \Omega} \bar{\t} \cdot \u \ ds,
 \label{eq:regularized}
\end{align}
where $g$ is a degradation function, which accounts for the loss of stiffness in the fracture.

Several choices of $g, w$ and $c_w$ are used in the literature, leading to various phase-field fracture formulations~\cite{amor2009regularized, pham2011gradient, burke2013adaptive, kuhn2015degradation, sargado2018high, bilgen2019detailed}.
In this work, we follow \cite{bourdin2000numerical, miehe2010thermodynamically} and employ $g(c):=(1-c)^2$, $w(c)=c^2$ and $c_w=2$, resulting in the widely used \textit{AT-2} phase-field fracture model proposed in~\cite{ambrosio1990approximation}.
Given these particular choices, it is possible to asymptotically show via $\Gamma$-convergence, that  the minimizer of~\cref{eq:regularized} tends to a minimizer of~\cref{eq:energy_not_reg}, as $l_s \rightarrow 0$, see \cite{giacomini2005ambrosio}.

In the next step, we reformulate the fracture surface integral from~\cref{eq:traction_forces}, into a computationally acceptable form, which does not include $\partial C$.
We follow~\cite{mikelic2014phase, mikeli2015, mikelic2015quasi} and employ Gauss' divergence theorem for extending the pressure $p$ to the entire domain, thus
\begin{align*}
\int_{\partial C} p \ \n \cdot \u \ ds  = \int_{\Omega} g(c) \nabla \cdot (p \  \u) \ d \Omega - \int_{\partial \Gamma_N} p \ \n \cdot \u \ ds.
\end{align*}
Here, the degradation function $g(c)$ ensures that the integration is performed only over the intact part of the domain $\Omega$.
Finally, the energy functional \cref{eq:energy_not_reg} can be recast into following form:
\begin{equation}
\begin{aligned}
    E(\u, c, p) &:=\int_{\Omega} g(c) \ \psi_e( \eps(\u)) +   \frac{\pazocal{G}_c}{c_w} \bigg( \frac{w(c)}{l_s} + l_s \left| \nabla c \right|^2  \bigg)  d\Omega  \\
    & - \int_{\Gamma_N} \bar{\t}_{\Omega} \cdot \u \ ds - \int_{\Omega} g(c) \nabla \cdot (p \u) \ d \Omega
 + \int_{\partial \Gamma_N} p \n \cdot \u \ ds,
\label{eq:pressurized}
\end{aligned}
\end{equation}
which can be employed in practical algorithms.

\subsection{Minimization problem}
The state of the system, defined by the displacement $\u$ and the phase-field $c$, is characterized at each loading step as  minimizer of the following minimization problem: \\
Find $(\u, c) \in \mathbf{V}_u^t \times V_c$, such that
\begin{align}
    (\u, c) \in \text{arg \ min} \  E(\u, c, p),
    \label{eq:min_problem_pf}
\end{align}
where the energy functional $E(\u, c, p)$ is as defined in~\cref{eq:pressurized}.
The admissible space for the displacement field is defined as  $\mathbf{V}_u^t := \{ \u \in \mathbf{H}^1(\Omega) \  | \  \u = \mathbf{g}^t \ \text{on} \   \Gamma_D \}$.
Here, $\mathbf{H}^1(\Omega):= [H^1(\Omega)]^d$, with $H^1$ denoting the standard Sobolev space of weakly differentiable functions in $L^2$ with one weak derivative also in $L^2$.
We remark that the definition of the space $\mathbf{V}_u^t$ incorporates the time-dependent Dirichlet boundary condition $\g^t$.
The admissible space for the phase-field is defined as a following convex cone:
\begin{align}
V_c := \{ c \in H^1(\Omega) :  c^{t-1}  \leq c \leq 1   \ \text{a.e. \ in}  \  \Omega  \},
\label{eq:space_pf}
\end{align}
where $c^{t-1} $ represents phase-field obtained in the previous time-step.
The box constraint $c^{t-1}  \leq c \leq 1 $ from~\cref{eq:space_pf} ensures the irreversibility condition and prevents the crack from self-healing.

We discretize our problem using the first order Lagrangian finite elements.
In the reminder of this work, we focus on the numerical solution of~\cref{eq:min_problem_pf}.
This task is numerically challenging and computationally demanding as we have to solve a large-scale, non-convex, constrained, ill-conditioned minimization problem for every loading time-step $t$.


\section{Multilevel trust-region method} \label{sec:rmtr}
In an abstract sense, the minimization problem \cref{eq:min_problem_pf} can be expressed in the following form:
\begin{equation}
\begin{aligned}
\min_{x \in \R^{n}} \ &f(\x), \\
\text{such that} \ &\x \in \pazocal{F},
\end{aligned}
\label{eq:min_problem}
\end{equation}
where $f: \R^{n} \rightarrow \R$ denotes the non-convex coupled energy functional~\cref{eq:pressurized} after  finite element discretization.
The solution vector $\x \in \R^{n}$ represents the combined displacement and phase-field coefficients.
The feasible set $\pazocal{F}:= \{ \x \in \R^{n} \ | \ \l \leq \x \}$ is defined such that irreversibility condition from~\cref{eq:space_pf} is satisfied.

For minimizing~\cref{eq:min_problem}, we employ the recursive multilevel trust-region method (RMTR)~\cite{Gratton2008recursive,gratton2008_inf,Gross2009}.
In particular, we employ the variant proposed in~\cite{kopanicakova20a}, which was specially designed to solve minimization problems arising from phase-field fracture simulations.
The use of the RMTR method is especially beneficial for large-scale simulations considered in this work since it combines the optimal complexity of multilevel methods with  the global convergence of the trust-region method.

By design, the RMTR employs a hierarchy of $L$ levels.
Each level $l$, where $l=1, \dots, L$, is associated with the minimization of some level dependent objective function $h^l: \R^{n^l} \rightarrow \R$.
Here, we assume that  $h^l$ is less costly to minimize than $h^{l+1}$ and that in particular  $n^{l+1} \geq n^{l}$.
The transfer of  data between subsequent levels of the multilevel hierarchy is achieved using three transfer operators.
The prolongation operator $\I^l: \R^{n^l} \rightarrow \R^{n^{l+1}}$ is used to interpolate the corrections from level $l$ to level $l+1$.
Its adjoint, the restriction operator $\Rev^l := (\I^l)^T$,  is used to transfer the dual quantities, such as gradients, to the next coarser level.
Following~\cite{Gross2009}, we additionally employ a projection operator $\P^l: \R^{n^{l+1}} \rightarrow \R^{n^{l}}$
for transferring iterates
to the next coarser level.

\subsection{RMTR Algorithm}
The RMTR algorithm is considered in its standard V-cycle form.
Through the following paragraphs, we use subscripts and superscripts to specify the iteration number and the given level respectively.
For instance, the symbol $\x^l_{i}$ denotes the solution vector on level $l$ during iteration $i$.

Each V-cycle consists of a downward and an upward phase.
The downward phase starts on the finest level, $l=L$, with an initial iterate $\x_0^L$ and passes through all levels of the multilevel hierarchy until the coarsest level, $l=1$, is reached.
On each level, the algorithm performs a pre-smoothing step in order to improve the current iterate $\x_{0}^l$.
This is done by minimizing the level dependent minimization problem, see~\cref{sec:ldmp}.
The minimization on a given level is performed only approximately, by employing $\mu_1$ iterations of the trust-region method.
The obtained approximate minimizer, $\x_{\mu_1}^L$,  is then used to initialize the solution vector on the next coarser level.
This is achieved using the projection operator defined above  as $\x_{0}^{L-1} := \P^{L-1} \x_{\mu_1}^L$.
We repeat this process recursively until the coarsest level is reached.

Once the coarsest level is reached, we again approximately minimize the level-dependent minimization problem in order to obtain an updated coarse grid iterate $\x_{\mu^1}^1$.
The minimization of on the coarsest level is usually carried out more accurately than on all other levels, for example by performing
additional trust-region steps.
After obtaining an updated iterate on the coarsest level,  $\x_{\mu^1}^1$, the RMTR algorithm initiates the upward phase of the V-cycle.
An upward phase is associated with the return to the finest level of the multilevel hierarchy while passing through all intermediate levels.
Starting on the coarsest level, we compute each coarse grid correction as the difference between the initial and final iterate on the given level, thus as $\x_{\mu^{l-1}}^{l-1} - \x_{0}^{l-1}$.
This coarse grid correction is then prolongated to the subsequent finer level, e.g. $\s_{\mu_1+1}^{l} :=  \I^{l-1}(\x_{\mu^{l-1}}^{l-1} - \x_{0}^{l-1})$.
As common in the trust-region algorithms, the quality of the prolongated coarse grid correction, $\s_{\mu_1+1}^{l}$, has to be assessed before it is accepted.
To this aim, we define a multilevel trust-region ratio as
\begin{align}
\rho^l := \frac{h^l(\x_{\mu_1}^l)  - h^l(\x_{\mu_1}^l  + \s_{\mu_1}^l)}{h^{l-1}(\x_{0}^{l-1})  - h^{l-1}(\x_{\mu^{l-1}}^{l-1})},
\label{eq:rho_ml}
\end{align}
where $\mu^l$ collectively denotes a sum of all iterations taken on a given level $l$.
The positive values of $\rho^l$ imply a decrease in the fine level objective function $h^l$, therefore it is safe to accept $\s_{\mu_1+1}^{l}$.
In contrast, small or negative values of $\rho^l$ suggest that there is no good agreement between fine and coarse level models, therefore $\s_{\mu_1+1}^{l}$ has to be rejected.

To this end, the RMTR algorithm performs $\mu_2$ smoothing steps in order to improve the current solution on a given level $l$.
This process is again repeated on every level of the multilevel hierarchy until we reach the finest level.
The outlined process is summarized in Algorithm~\ref{alg:rmtr}.

\subsection{Level dependent minimization problems}
\label{sec:ldmp}
On each level of the multilevel hierarchy, the RMTR method minimizes the following level dependent minimization problem:
\begin{equation}
\begin{aligned}
\min_{\s^l \in \R^{n^l}} h^l(\x^l + \s^{l}),  \\
\text{subject to} \  \x^l + \s^{l} \in \pazocal{F}^l,
\end{aligned}
\label{eq:level_dep_min_problem}
\end{equation}
where $h^l$ and  $\pazocal{F}^l$ denote the level-dependent objective function and feasible set respectively.
Here, we follow \cite{kopanicakova20a}, and define $h^l$ as
\begin{align}
h^l(\x^l +  \s^{l}) :=
\begin{cases}
\tilde{f}^l(\x^l) + \langle \delta \g,  \s^{l} \rangle  + 0.5 \langle  \s^{l}, \delta \H  \s^{l}    \rangle, \ &\text{if} \ l < L, \\
f(\x^l), & \text{if} \ l = L.
\end{cases}
\label{eq:second_order_consistency}
\end{align}
Thus, on the finest level $h^l$ is identical with the target objective function $f$.
On all other levels, $h^l$ is defined using modified energy functional $\tilde{f}^l(\x^l)$, which allows us to combine
the fine level description of the fractures with the coarse level discretization, as in~\cite{kopanicakova20a}.
The terms $\delta \g \in \R^{n^l}$ and $\delta \H \in \R^{n^l \times n^l}$ from~\cref{eq:second_order_consistency},  defined as
\begin{equation}
\begin{aligned}
\delta \g := &\Rev^l \nabla h^{l+1}(\x_{\mu_1}^{l+1}) - \nabla  \tilde{f}^l(\x^l_0),\\
\delta \H := &\Rev^l \nabla^2 h^{l+1}(\x_{\mu_1}^{l+1}) \  \I^l - \nabla^2  \tilde{f}^l(\x^l_0),
\end{aligned}
\label{eq:delta_terms}
\end{equation}
ensure that the first and second order behavior of the $h^l$ and $h^{l+1}$ is similar in the neighborhood of $\x_0^l$ and $\x_{\mu_1}^{l+1}$.

The role of level-dependent feasible set $\pazocal{F}^l$ is two-fold.
On the one hand, it ensures that the iterates produced by the RMTR method satisfy the variable bounds.
On the other hand, the definition of  $\pazocal{F}^l$ also controls the size of all corrections taken on a given level $l$, which is necessary in order to ensure global convergence \cite{gratton2008_inf}.
The rigorous details about how to construct $\pazocal{F}^l$ can be found in \cite{gratton2008_inf,Kornhuber1994,kopanicakova20a}.

\begin{algorithm}
\caption{V-cycle of RMTR ( $l, \  \x^l_{0},  \  \Rev \g,  \  \Rev \H, \  \pazocal{F}^l,  \  \Delta^l_{0}$)}
\label{alg:rmtr}
\begin{algorithmic}
\Require{$l \in \R, \x^l_{0} \in \R^{n^l},  \Rev \g \in \R^{n^l}, \Rev \H \in \R^{n^l \times n^l}, \pazocal{F}^l,  \Delta^l_{0} \in \R$}
\Constants { $\mu_1, \mu_2 \in \N$}
\State Generate $h^l$ by means of \cref{eq:second_order_consistency} using $\Rev \g, \Rev \H$
\State $ [\x_{\mu_1}^l, \  \Delta_{\mu_1}^l] = \text{ Nonlinear}\_\text{solve}(h^l,  \ \x_{0}^l, \  \pazocal{F}^l,  \  \Delta_{0}^l, \  \mu_1 )$
\State Generate coarse level feasible set $\pazocal{F}^{l-1}$
\State Evaluate $\Rev \g=\Rev^{l-1} \nabla h^l(\x_{\mu_1}^l)$, $\Rev\H=(\Rev^{l-1})^T \nabla^2 h^l(\x_{\mu_1}^l)  \I^{l-1}$
\If{$l == 2$}
\State Generate $h^{l-1}$ by means of  \cref{eq:second_order_consistency} using $\Rev \g, \Rev \H$
\State $ [\x_{*}^{l-1}] = \text{ Nonlinear}\_\text{solve}(h^{l-1},  \ \P^{l-1} \x_{\mu_1}^l, \  \pazocal{F}^{l-1},  \  \Delta_{\mu_1}^l, \  \mu^1 )$
\Else
\State  $ [\x_{*}^{l-1}] = \text{RMTR}(l-1,  \ \P^{l-1} \x_{\mu_1}^l, \Rev \g,  \ \Rev \H, \  \pazocal{F}^{l-1}, \  \Delta_{\mu_1}^l)$
\EndIf
\State $\s^l = \I^{l}(\x_{*}^{l-1} - \P^{l-1} \x_{\mu_1}^l)$
\State Evaluate $\rho^l$ by means of \cref{eq:rho_ml}
\State $[\x_{\mu_1 +1}^l, \  \Delta_{\mu_1+1}^l] = \text{Convergence\_control}(\rho^l, \ \x_{\mu_1}^l, \  \s^l, \  \Delta_{\mu_1}^l)    $
\State
\State $ [\x_{*}^l, \  \Delta_{*}^l] = \text{ Nonlinear}\_\text{solve}(h^l,  \ \x_{\mu_1 +1}^l, \  \pazocal{F}^l,  \  \Delta_{\mu_1+1}^l, \  \mu_2 )$
\State
\Return  $\x_{*}^l, \Delta_{*}^{l}$
\end{algorithmic}
\end{algorithm}

\subsubsection{Smoothing and coarse grid solve (trust-region method)}
On each level $l$, the RMTR method requires an approximate solution of level-dependent minimization problem~\cref{eq:level_dep_min_problem}.
To this aim, we employ a trust-region method~\cite{conn2000trust},  shown in Algorithm~\ref{alg:tr}.
\begin{algorithm}
\caption{Nonlinear\_solve( $h, \  \x_{0},  \ \pazocal{F},  \  \Delta_{0},  \ i_{\text{max}}$)}
\label{alg:tr}
\begin{algorithmic}
\Require{$h:\R^n \rightarrow \R,\ \x_{0} \in \R^n, \  \pazocal{F}, \   \Delta_{0} \in \R, \  i_{\text{max}} \in \N $}
\For{$i=0, \dots, i_{\text{max}}$}
\State
\State $ \underset{\s_i \in \R^n}{\text{min}}  \ m_i(\s_i) =h(\x_i) + \langle \nabla h(\x_i), \s_i  \rangle  + 0.5 \langle  \s_i, \nabla^2 h(\x_i) \s_i \rangle $
\State $\text{subject to} \  \x_i + \s_i \in \pazocal{B}_i \cap \pazocal{F} $
\State
\State Evaluate $\rho$ by means of \cref{eq:rho_tr}
\State $[\x_{i +1}, \  \Delta_{i+1}] = \text{Convergence\_control}(\rho, \ \x_{i}, \  \s_{i}, \  \Delta_{i})    $
\EndFor
\State
\Return  $\x_{i +1},\ \Delta_{i+1}$
\end{algorithmic}
\end{algorithm}
The following exposition omits using superscript related to a given level $l$, as all quantities are considered to be on the same level.
At each iteration $i$, the trust-region method approximates the objective function $h$ by quadratic model $m_i$, defined around current iterate $\x_i$.
The model $m_i$ is considered to be an adequate representation of $h$ only in a certain region, called the trust-region $\pazocal{B}_i := \{ \x_i + \s \in \R^n  \ | \  \| \s \|_{\infty} \leq \Delta_i \}$,  defined by the trust region radius $\Delta_i > 0$.
The search direction $\s_i$ is then determined by solving following trust-region sub-problem:
\begin{align}
\underset{\s_i \in \R^n}{\text{min}} \ m_i(\s_i ) &:= h(\x_i) + \langle \nabla h(\x_i), \s_i \rangle + \frac{1}{2} \langle \s_i,  \nabla^2 h(\x_i) \  \s_i \rangle, \nonumber  \\
&\text{such that}  \ \ \x_i + \s_i \in  \pazocal{F}, \label{eq:model_qp}  \\
& \ \  \| \s_i \|_{\infty} \leq \Delta_i. \nonumber
\end{align}
The first constraint in~\cref{eq:model_qp} ensures the feasibility of the iterates through the solution process, while the second constraint controls the size of the search direction $\s_i$.
Before the minimizer of~\cref{eq:model_qp}, search direction $\s_i$, is used to update the current iterate $\x_i$, we need to assess its quality.
The convergence control, Algorithm~\ref{alg:conv_control},  is performed using the trust-region ratio
\begin{align}
\rho_i =  \frac{h(\x_i) - h(\x_i + \s_i)}{m_i(\bm{0}) - m_i(\s_i) },
\label{eq:rho_tr}
\end{align}
which describes the agreement between the actual reduction in the objective function and the predicted reduction obtained by the quadratic model $m_i$.
The value of $\rho_i$ close to unity indicates good agreement between $f_i$ and the model $m_i$.
Hence, it is safe to accept $\s_i$, i.e. $\x_{i+1} = \x_i + \s_i$, and expand the trust-region radius.
In contrast, if the value of $\rho_i$ is negative or close to zero, we must reject $\s_i$, i.e. $\x_{i+1}=\x_i$, and shrink the trust region.

\begin{algorithm}
\caption{Convergence\_control( $\rho, \  \x_{i},  \ \s_i,  \  \Delta_{i}$)}
\label{alg:conv_control}
\begin{algorithmic}
\Require{$\rho \in \R,\ \x_{i} \in \R^n, \s_{i} \in \R^n,  \  \Delta_{i} \in \R $}
\Constants { $\eta_1, \eta_2, \gamma_1, \gamma_2 \in \R$, where $0 < \eta_1 \leq \eta_2 \leq 1$ and $0 < \gamma_1 \leq 1 \leq \gamma_2$}
\end{algorithmic}
\hfill
\begin{minipage}[t]{0.4\columnwidth}
\begin{algorithmic}
\If{$\rho > \eta_1$}
\State  $\x_{i+1} = \x_{i} + \s_i$
\Else
\State   $\x_{i+1} = \x_{i}$
\EndIf
\State
\Return  $\x_{i +1},\ \Delta_{i+1}$
\end{algorithmic}
\end{minipage}
\hfill
\begin{minipage}[t]{0.55\columnwidth}
\begin{algorithmic}
\State
\State  $\Delta_{ i+1} =
\begin{cases}
\gamma_1  \Delta_{i}, & \rho < \eta_1 \\
\Delta_{i} , & \rho \in  [ \eta_1, \eta_2]   \\
 \gamma_2 \Delta_{i}, & \rho >\eta_2
\end{cases}
$
\State
\end{algorithmic}
\end{minipage}
\end{algorithm}

\paragraph{Solution of trust-region subproblem} \label{sec:quadprog}
Each iteration of the TR method, Algorithm \ref{alg:tr}, requires solution of constrained quadratic minimization (QP) problem \cref{eq:model_qp}.
The arising QP problems can be solved approximately, as long as the obtained minimizer satisfies the so-called sufficient decrease condition~\cite{conn2000trust}.
Our choice of QP solver varies for different levels of the multilevel hierarchy.
In particular, on the coarsest level, we minimize \cref{eq:model_qp} using Modified Proportioning with Reduced Gradient Projection (MPRGP) method~\cite{dostal2016mprgp}.
On all the other levels, we employ only few steps of the projected Gauss-Seidel (PGS) method, as it is known to have good smoothing properties \cite{briggs2000multigrid, hackbusch2013multi}.
Since the Gauss-Seidel method is naturally a sequential algorithm, we employ its parallel variant, the hybrid Jacobi projected-Gauss-Seidel (HJPGS) method~\cite{adams2003parallel}.
More specifically, we use the symmetric version of the HJPGS where both forward and back substitution are performed.


\section{Implementation} \label{sec:implementation}
\subsection{Workflow}  \label{sec:worflow}
\begin{figure}
    \centering\footnotesize
    \includegraphics[width=0.98\linewidth]{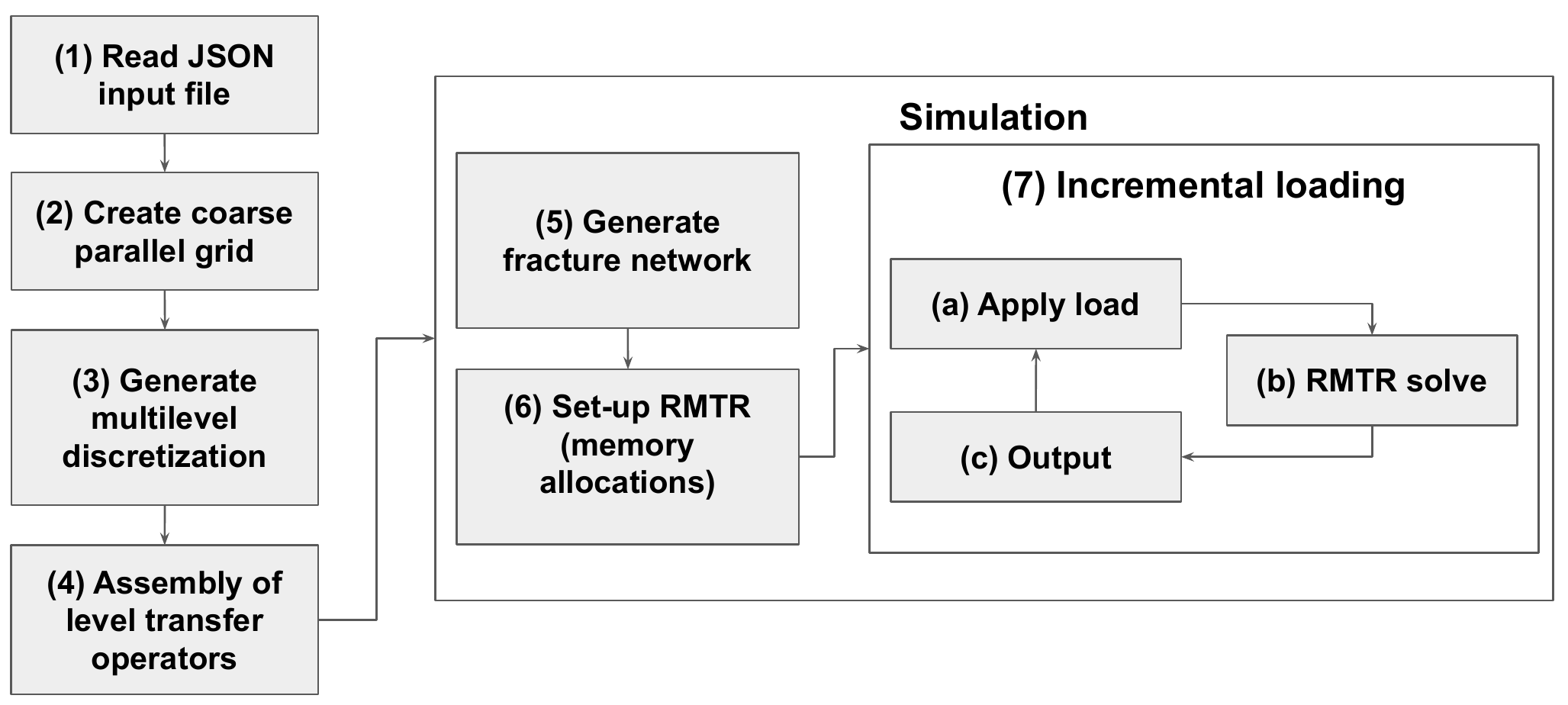}
    \caption{Simulation workflow.}
    \label{fig:workflow}
\end{figure}

The workflow of our simulation, illustrated in \Cref{fig:workflow}, consists of the following steps:
\begin{enumerate}
    \item reading the simulation parameters from a JSON (JavaScript Object Notation) file~\cite{json};
    \item creating a distributed structured coarse grid;
    \item constructing the multilevel hierarchy by uniform refinement and level dependent minimization problems~\cref{eq:level_dep_min_problem} using linear finite element spaces;
    \item assembling the transfer operators for exchanging the discrete fields between subsequent levels of the hierarchy;
    \item generating the stochastic fracture network;
    \item setting-up all the levels of RMTR by allocating all the necessary buffers (including temporaries) and initializing the QP solvers;
    \item performing incremental loading, where we repeatedly
    \begin{enumerate}
        \item increase loading conditions;
        \item minimize~\cref{eq:min_problem_pf} using RMTR for updating the solution;
        \item export the solution to disk.
    \end{enumerate}
\end{enumerate}
While steps 2-4 mainly use PETSc-DMDA facilities directly, all the solvers are realized by exclusively using the Utopia front-end.

\subsection{RMTR and HPC} \label{sec:perfdetails}
Most of the computational time is invested in computing the Hessian and in solving the trust region sub-problems~\cref{eq:model_qp}.
Therefore, we describe some of the details that mainly define the resulting performance of our implementation of these two aspects.

\subsubsection{Hessian assembly}
Given our RMTR setup, where $\mu_1=\mu_2=1$, the method requires three Hessian evaluations per level during each V-cycle.
We can decrease the amount of assembly calls by incorporating the Hessian lagging strategies into our implementation.
In particular, we evaluate the $\delta \H$ term from~\cref{eq:delta_terms} restricting the Hessian evaluated during the pre-smoothing step.
Additionally, we skip the Hessian evaluation while performing post-smoothing steps, and employ the Hessian assembled during the pre-smoothing step.
In this way, the RMTR method requires only one Hessian evaluation on each level of multilevel hierarchy per V-cycle.
We note, that this modifications slightly worsen the convergence rate of the RMTR method, but offer $20-30\%$ speed-up in terms of the computational time.

The assembly of the Hessian involves performing numerical quadrature for each element at every nonlinear iteration, and has to be performed on two-dimensional and three-dimensional geometries with different element types, which also require different quadrature rules.
For dealing with this variety of inputs and preserve performance, we use compile time dimensions (by means of template parameters) for all loop ranges for allowing the compiler to perform aggressive optimizations automatically.
We are exclusively using structured grids which allows us to pre-compute several quantities for one element and reuse them for all elements.
In fact, we pre-compute all linear components associated with the model, which include all test-space related quantities such as shape-functions, gradients, strains, principal strains and stresses, and geometric quantities such as Jacobian matrices and determinants.
A similar idea could also be used for isotropic adaptive octree grids while maintaining a low memory footprint where the element volumetric uniform-scaling parameter would be used together with the pre-computed quantities when performing quadrature.
However, due to fact that the model is nonlinear, we are required to compute many quantities based on the current solution.
To this purpose our quadrature routines exploit inline expression-template evaluation for small tensor in order to keep the code readable and avoid creating temporaries.

\subsubsection{Constrained QP-solvers}
After the computation of the Hessian, the HJPGS (introduced in \cref{sec:quadprog}) is the most computationally expensive part of our code.
For mitigating this issue we apply few minute measures.
First, we copy the local diagonal block of the Hessian data and separate diagonal and off-diagonal entries and store them in different arrays.
This is mainly done in order to avoid checking if the current row is equal to the current column. We achieve around 2x speed-up compared to the version without a copy.
Second, we can perform local iterations of the smoother without synchronization in order to reduce the ratio between computation and communication.


\begin{figure} \centering\footnotesize
   \includegraphics[width=0.43\linewidth]{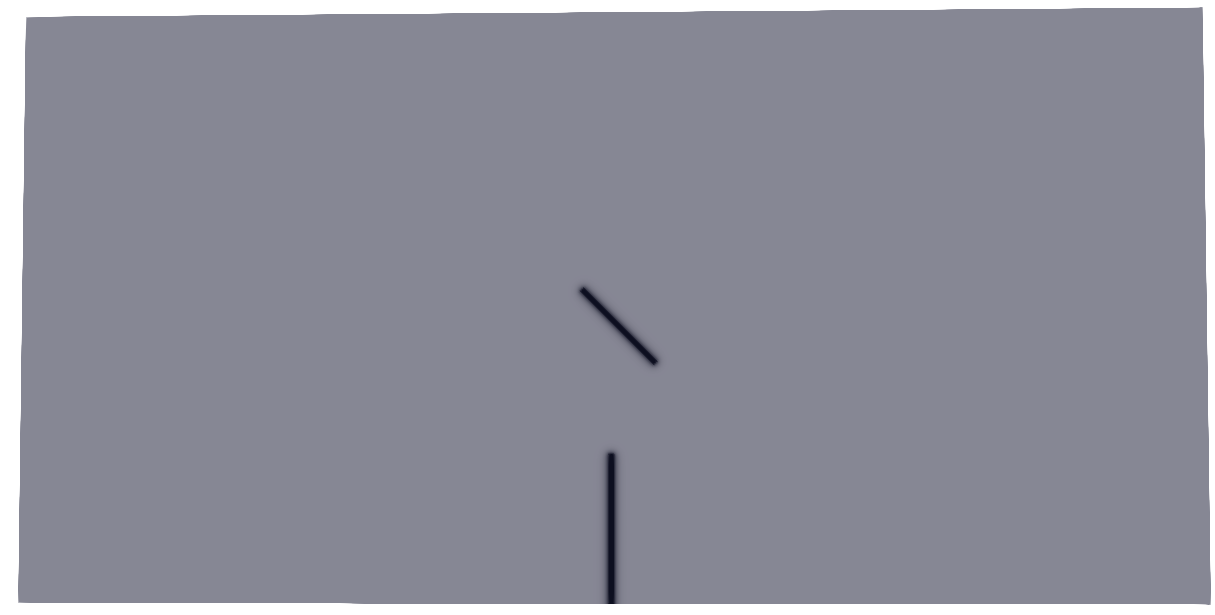}\hfill
   \includegraphics[width=0.43\linewidth]{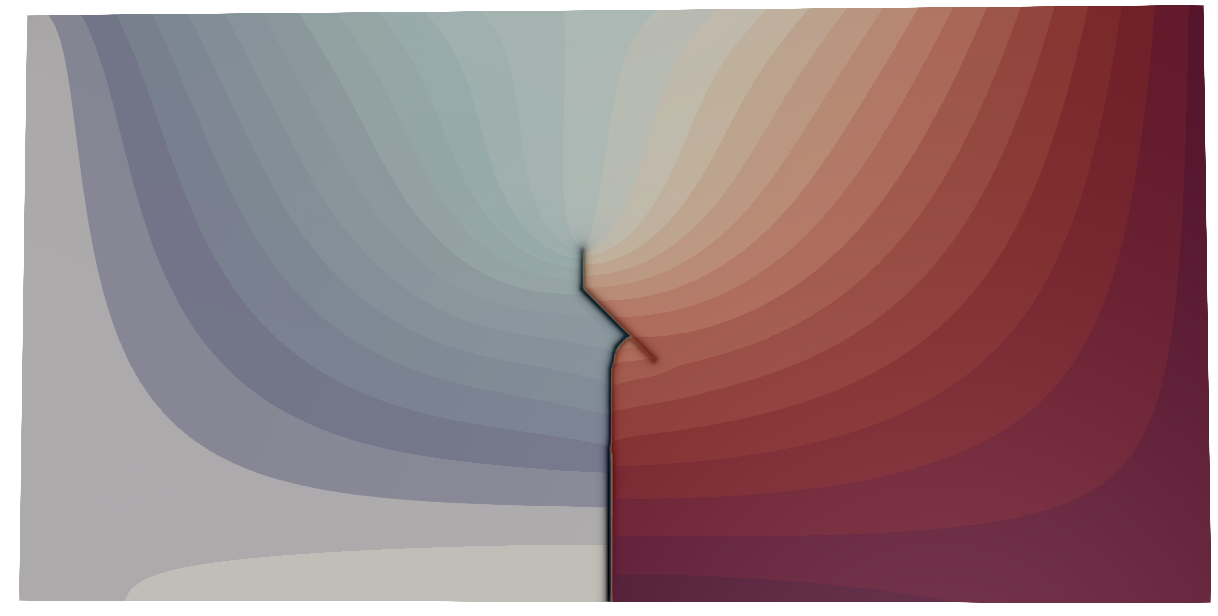} \hfill
   \includegraphics[width=0.067\linewidth]{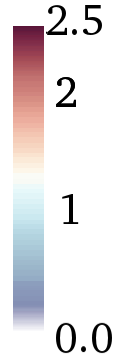}
   \caption{Two-dimensional simulation with $2$ fractures and $949\,227$  degrees of freedom. Color represents displacement field in mm.}
   \label{fig:AT}
 \end{figure}

\section{Numerical experiments} \label{sec:experiments}
For initializing $c$ to its transitional state from intact to broken, we check if the nodal position $\x$ lies inside of a parametric fracture description, then we mark the related parts of the domain as broken.
This is done by setting the nodal coefficient for the phase-field variable to be equal to $1$.
Otherwise, we mark the material as intact by prescribing the nodal value of the phase-field to be equal to $0$.
We first validate our code using experimental measurements, then we consider more complicated scenarios inspired by hydraulic simulations performed in enhanced geothermal systems.
The value of the length-scale parameter $l_s$ is set up as $l_s=2h$, where $h$ denotes the mesh size, for all presented numerical examples.

The main output data of the experiments can be downloaded from the Zenodo online repository~\cite{patrick_zulian_2020_3760411}.

\subsection{Validation}
\label{sec:avalidation}
We consider two initial cracks inserted in an asphalt specimen. The initial crack length is set equal to $a = 5\,\text{mm}$, the initial width is set equal to $w_0=0.2\,\text{mm}$, whereas the relative positions of the two cracks is defined such that they comprise an angle equal to $45\,^\circ $.
The background matrix is a two-dimensional rectangle with height equal to $20\,\text{mm}$ and width equal to $40\,\text{mm}$.
The Lam\'e parameters of the asphalt are $\mu= 2.23\,\text{N/mm}^2$ and $\lambda = 3.35\,\text{N/mm}^2$, whereas the fracture energy is set equal to $\pazocal{G}_c=0.270\,\text{N/mm}$ in agreement with ~\cite{hou2014computational}.
Concerning the boundary conditions, we fix the left side of the rectangle whereas an incremental displacement is applied on the right side and defined as $u(t)=u_0+\Delta t u_0$ with $u_0=3.0\,\text{mm}$ and $\Delta t =0.01\,\text{s}$.

In \Cref{fig:AT} we show the initial and final configuration, where the two fractures interact with each other. Here, the mean displacement reached on the right side of the sample, $u=2.366\,\text{mm}$, corresponds to a critical load $\sigma_{c}^{n}=0.343\,\text{MPa}$, in good agreement with the experimental result $\sigma_{c}^{e}=0.30\,\text{MPa}$ reported in~\cite{hou2015fracture}.

  \begin{figure}
   \includegraphics[width=0.99\linewidth]{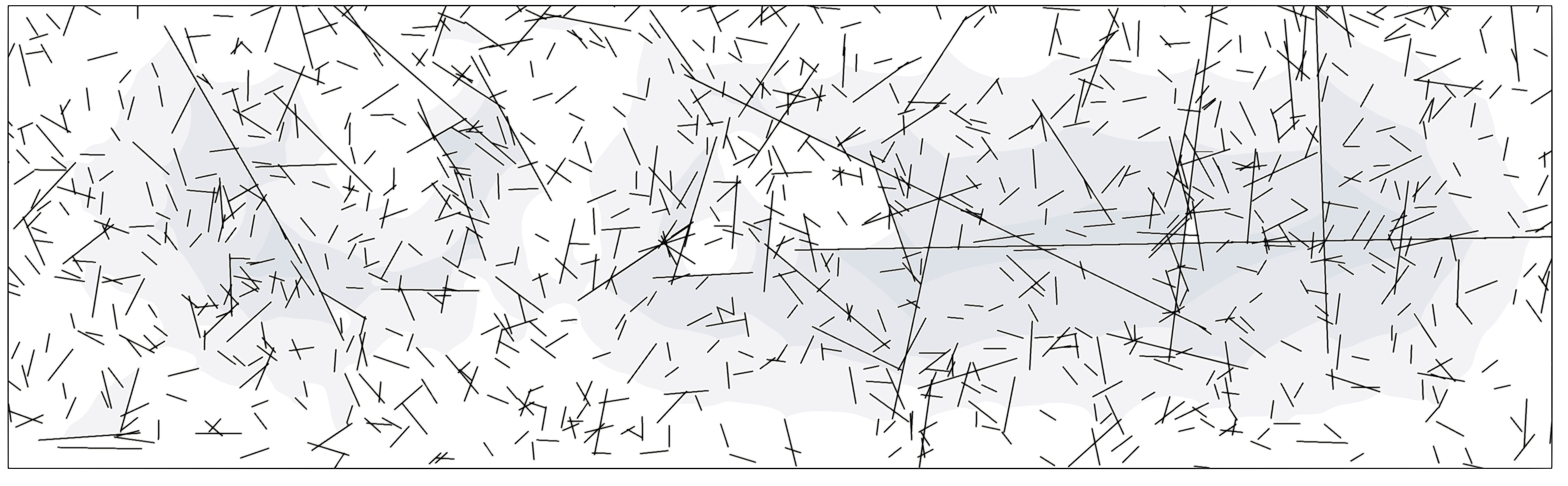} \\
   \includegraphics[width=0.99\linewidth]{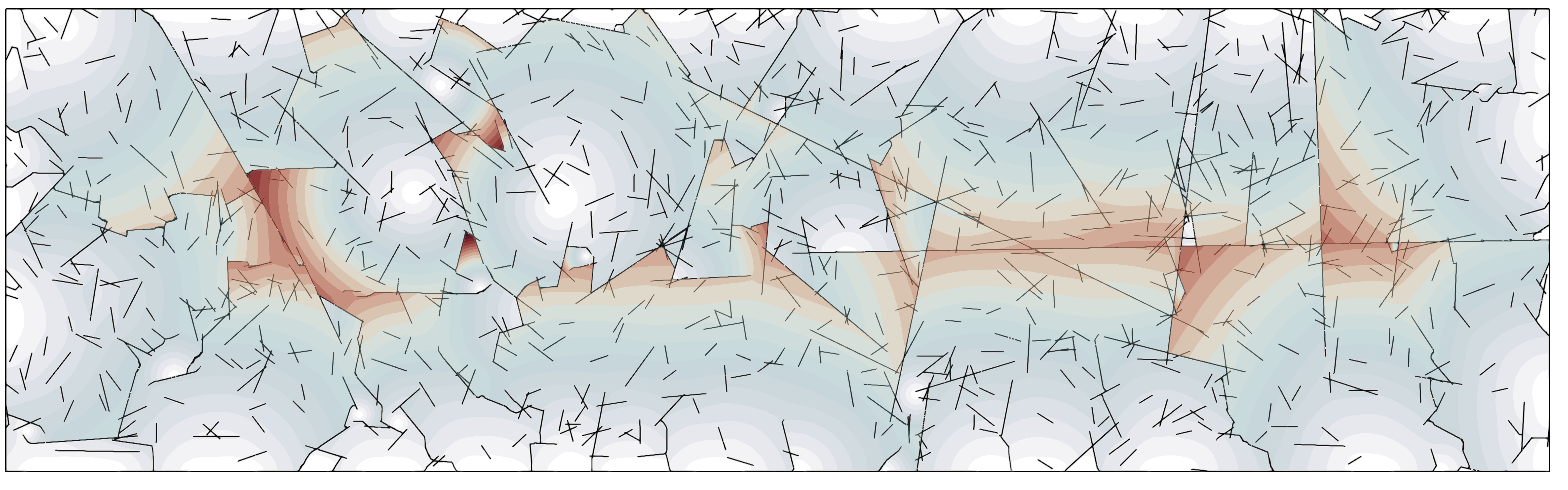}
   \caption{Two-dimensional simulation with $1000$ fractures and $13\, 565\, 475$ degrees of freedom. The colored overlay represents the displacement magnitude [0, 1.5] mm from transparent blue to opaque red.
   Left: initial fracture network. Right: final configuration.}
   \label{fig:teaser}
 \end{figure}

\subsection{Geoscience application}
\label{sec:application}
The presented numerical experiments consider pressure induced fracture propagation of realistic stochastic fracture networks in two-dimensional and three-dimensional scenarios. We generate the pre-existing fracture networks using a two-stage process.
In the following, we give details about the procedure used to generate a one-dimensional fracture network embedded in a two-dimensional background matrix (\Cref{fig:teaser}).
First, we describe each fracture as a one-dimensional object with a randomly assigned hypo-center, orientation, and length.
In particular, we employ a uniform distribution to place the hypo-centers over the entire domain and assign their orientation to a value between $-80 ^\circ $ and $80 ^\circ $. The fracture length is drawn from a scale-invariant power-law distribution~\cite{de2001hydraulic}.

In the second stage,  each fracture is regularized through a volumetric representation
with artificial width $w$ proportional to the mesh size $h$, where $w=2\,h$.
Hence, the resulting fracture networks consist of \textit{smooth rectangles} randomly embedded in the surrounding matrix.

It is worth pointing out that a similar procedure is also used in three-dimensional simulations, where the fracture network consists of \textit{smooth parallelepipeds} randomly distributed inside the three-dimensional background material.
The fracture network represents the initial datum for the phase-field parameter which evolves during the simulation depending on the prescribed pressure and boundary conditions.

\subsubsection{Three-dimensional scenarios}
\begin{figure*} \centering\footnotesize
   \includegraphics[width=0.2\linewidth]{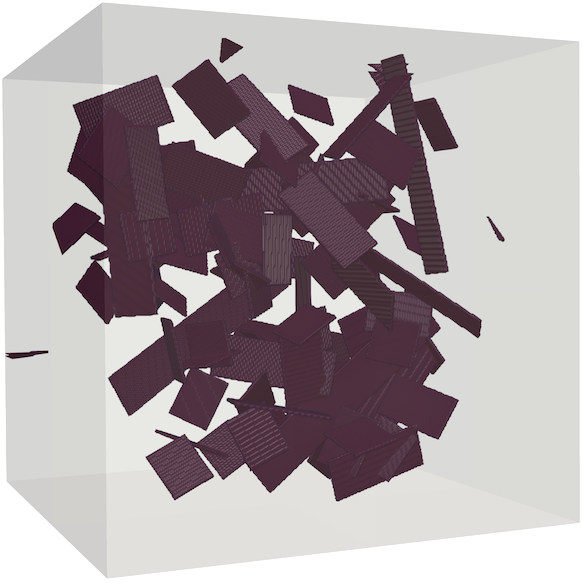} \hfill
   \includegraphics[width=0.2\linewidth]{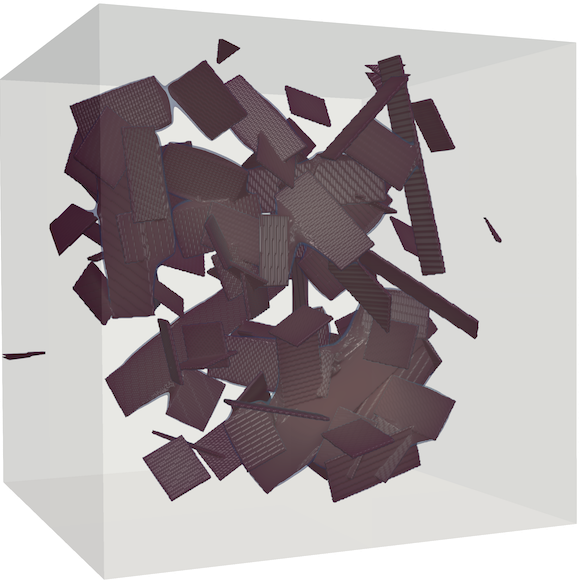} \hfill
   \includegraphics[width=0.2\linewidth]{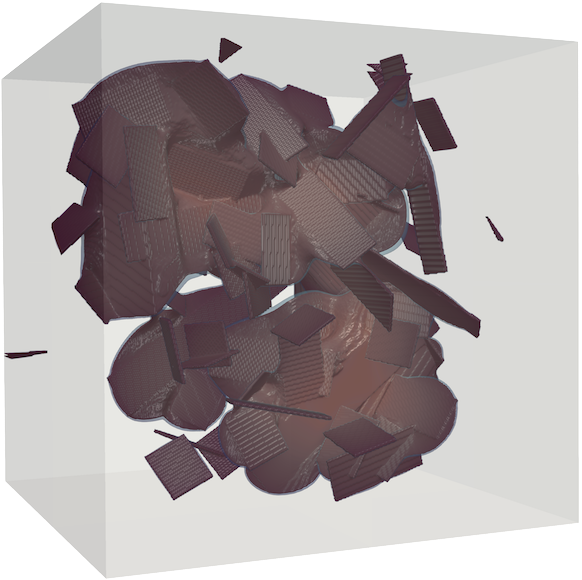} \hfill
   \includegraphics[width=0.2\linewidth]{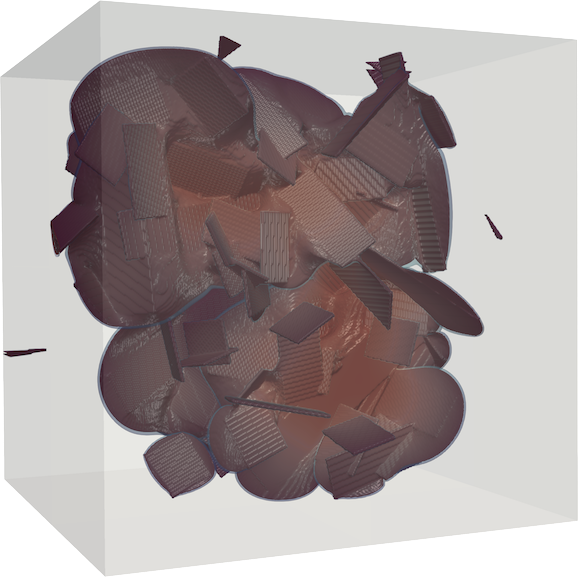}
   \caption{Four loading steps of a three-dimensional simulation with $100$ randomly distributed fractures and $242\,793\,828$ degrees of freedom (number of levels is 4).
   The fracture iso-surface is displayed for $c=0.9$.
   The colored transparent overlay represents the displacement magnitude [$0, 0.0032$] mm from blue to red.
   Snapshots taken at different times $t \in \{0, 0.75, 0.8, 0.9 \}$ s.}
   \label{fig:sim3D}
 \end{figure*}

We consider a network of pneumatic fractures embedded in a three-dimensional background matrix consisting of a cube with size $1 \times 1 \times 1\text{mm}$. The initial set-up of the simulation takes into account $100$ randomly distributed fractures as shown in \Cref{fig:sim3D}.
The critical energy release rate is set equal to $\pazocal{G}_c = 1 \text{N/mm}$ whereas the Lam\`e parameters are set equal to $\lambda = 100\,000 \text{N/mm}^2$ and $\mu = 100\,000 \text{N/mm}^2$, respectively, and describe the mechanical response of granite material~\cite{yu2018calculation}.
 We apply zero Dirichlet boundary conditions for the displacement field on all sides of the domain.  A pressure load is linearly increased at each loading step and defined as $ p(t) =p_0 + \Delta t p_c$, with $p_0= 0.010 \text{N/mm}^2$, $\Delta t=  0.05\,\text{s}$ and $p_c=1.0 \text{N/mm}^2$.
 The number of degrees of freedom of the simulation is $242\,793\,828$.
 \Cref{fig:sim3D} depicts the evolution of the fracture network together with the displacement field.
 One may observe interesting crack patterns and interacting fractures inside the computational domain.

\section{Performance and scaling}
\label{sec:perf}
All experiments have been performed at the Swiss National Supercomputing Centre (CSCS) with the Piz Daint super-computer on
XC50 compute nodes\footnote{A XC50 node consists of one Intel \textsuperscript{\textregistered} Xeon \textsuperscript{\textregistered} E5-2690 v3 @2.60 GHz (12 cores, 64 GB RAM)}. Every experiment uses all 12 cores (without hyperthreading) of a node. Thus, an experiment running on 4 nodes is in fact running with $12\times4=48$ MPI processes. We focus our scaling tests on measuring the cost of the nonlinear multilevel operator, therefore we set all convergence tolerances to zero and limit the number of maximum iterations. We observe that PGS converges, for constant grid size, faster on fewer processes (\Cref{tab:slowdown}). This effect is mitigated and we compare pure scaling performance excluding  effects where a solver converges with fewer iterations.

We traced the code to understand its parallel behaviour using mpiP \cite{vetter2005mpip} on XC40 compute nodes\footnote{A XC40 node consists of two Intel \textsuperscript{\textregistered} Xeon \textsuperscript{\textregistered} E5-2695 v4 @2.10 GHz (2 x 18 cores, 64/128 GB RAM)}.
Subsequently we have run a test with a grid of 40x40x40, 4 levels totalling  $122.6$ Million dofs over 1152 MPI tasks on the finest grid.
Among all MPI calls, $79\%$ of the MPI time is due to three calls: AllReduce, Iprobe and Test.
AllReduce calls are the most demanding. The heaviest are called in the calculation of the norms in the QP solvers, they count for $50\%$ of the MPI time and the $16\%$ of the overall application's time. Following the reductions, Iprobe and Test calls, which are called by the matrix assembly, are noticeable for roughly $7\%$ of the MPI time.

\subsection{Strong scaling}
We conducted two strong scaling experiments one \textit{small} with a coarse grid of $25 \times 25 \times 25$, 4 levels  totalling $28.7$ million dofs on the finest grid  and one \textit{big} experiment with a coarse grid of $50 \times 50 \times 50$, 4 levels totalling $242.7$ million dofs on the finest grid. The small experiment was run on $4$, $5$, $6$, $7$, $8$, $12$, $16$, $20$, $24$, $28$, $32$ nodes, the big one on $40$, $48$, $56$, $64$, $80$, $96$, $112$, $128$, $160$, $192$, $224$, $256$. The minimal number of nodes was chosen in such a way that the experiment fits into the node's RAM. In \Cref{fig:scaling}(a,b) we see the parallel efficiency that is defined as $e=\frac{T_b n_b}{T_n n}$ with $T_b$ being the base experiment's runtime ($n_b=4$ and $n_b=40$ nodes respectively) and  $T_n$ being the experiment's runtime on $n$ nodes. We can see that the parallel efficiency oscillates depending on the number of nodes. We assume that this is due to slight imbalances which appear to have sometimes a bigger effect on the total runtime than with the same coarse grid size but a different node count.

\begin{table}
\begin{center}
 \begin{tabular}{|c| c| c| c|c|c|c|} 
 \hline
 $\#$ nodes & $4$ & $8$ & $16$  & $32$ \\ 
 \hline\hline
 $\#$ V-cycles & $126$ & $135$ & $147$ & $154$ \\ 
 slow down & $0 \%$ & $10 \%$ & $16 \%$ & $22\%$ \\ 
 \hline
\end{tabular}
\end{center}
\caption{
Effects of PGS convergence on performance.
Average number of V-cycles over all time-steps as a function of number of nodes.
Three dimensional experiment performed with $100$ fractures, $28.7$ mil. dofs on the fine level, RMTR setup with four levels.}
\label{tab:slowdown}
\end{table}

\subsection{Weak scaling}
\begin{figure*}
 \begin{tikzpicture}
    \begin{axis}[
    scale=0.39,
    grid=major,
    ymin=0.9,
    ymax=1.3,
    xmode = normal,
    ymode=normal,
    xlabel={\# nodes},
    tick label style={font=\footnotesize},
    label style={font=\footnotesize},
    legend style={font=\footnotesize},
    ylabel=Parallel efficiency
    ]
    \addplot[color = myblue, very thick, mark=x] table [x=nodes, y=main_efficiency_total, col sep=comma] {strong_scaling_25x25x25.csv};
    \end{axis}
  \end{tikzpicture}
  \hfill
  \begin{tikzpicture}
    \begin{axis}[
    scale=0.39,
    grid=major,
    ymin=0.5,
    ymax=1.05,
    xmode = normal,
    ymode=normal,
    xlabel={\# nodes},
    ylabel=Parallel efficiency,
    tick label style={font=\footnotesize},
    label style={font=\footnotesize},
    legend style={font=\footnotesize},
    ]
    \addplot[color = myblue, very thick,mark=x] table [x=nodes, y=main_efficiency_total, col sep=comma] {strong_scaling_50x50x50.csv};
    \end{axis}
  \end{tikzpicture}
  \hfill
  \begin{tikzpicture}
    \begin{axis}[
    scale=0.39,
    grid=major,
    xmin=1,
    xmax=1024,
    ymin=0.5,
    ymax=1.1,
    xmode = log,
    ymode=normal,
    xlabel={\# nodes},
    ylabel=Parallel efficiency,
    tick label style={font=\footnotesize},
    label style={font=\footnotesize},
    legend style={font=\footnotesize},
    ]
    \addplot[color = myblue, very thick, mark=x] table [x=nodes, y=main_efficiency_total, col sep=comma] {weak_scaling.csv};
    \addplot[color = myred, very thick, mark=0] table [x=nodes, y=main_efficiency_total_corrected, col sep=comma] {weak_scaling.csv};
    \end{axis}
  \end{tikzpicture}
  \caption{a) Strong scaling performed with  $28.7$ mil. dofs on the fine level, RMTR setup with four levels.
  b) Strong scaling performed with $242.7$ million dofs  on the fine level, RMTR setup with four levels.
  c) Weak scaling with a lower (blue) and upper (red) estimate of parallel efficiency.  }
  \label{fig:scaling}
\end{figure*}
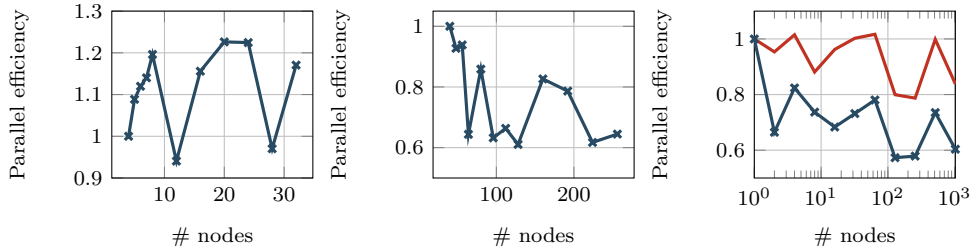

For weak scaling we have set up the experiment with a coarse grid of $10 \times 10 \times 10$ on a single node and incremented then by doubling the nodes and adapting the dimensions to have a similar number of dofs on the coarse grid. Experiments with a cube number of nodes are exact in the sense that the work per node on the coarse grid is exactly the same as for the base experiment on one node. In \Cref{fig:scaling}(c) we can see the results for the parallel efficiency defined as $e=T_b/T_n$ with $T_b$ being the base experiment's runtime ($10 \times 10 \times 10$ on one node) and $T_n$ being the runtime of the experiment on $n$ nodes. Additionally we have a red line which gives us an upper estimate of the parallel efficiency. It is a ``corrected'' value where we multiply $e$ with a constant $c=\frac{N}{N_bn}$ with $N$ being the number of dofs on the finest grid and $N_b$ being the number of dofs on the finest grid for the experiment on one node. This correction factor is larger than 1, because doubling each dimesion on the coarse grid will increase the number of dofs by a factor larger than 8 on the finest grid. For a setup with 4 levels the number of dofs on the finest grid in $x$-direction is $8N_x-7$, similarly in $y$ and $z$-direction, which results in larger multiplication factor on the finest level than the multiplication factor on the coarse level.

\section{Conclusion} \label{sec:conclusion}
We presented the first open-source code for numerical modelling of large scale phase-field fracture simulations using the RMTR method.
Our implementation of the phase-field fracture model employs an expression template based assembler designed for structured grids and 2D/3D tensor-product finite elements.
Our implementation of the RMTR method with its different components, such as the quadratic programming solvers, provided in the Utopia software library can deal with non-convex and geometrically complex problems in an efficient and scalable way.
Every aspect of the code has been first optimized for single core CPU performance, then improved for MPI based parallelism.

All the numerical examples show the capabilities of our simulation framework and its suitability for large scale geoscience applications, such as hydraulic fracturing of complex fracture networks.
To this end, our studies show the parallel performance by analyzing strong and weak scaling properties to the limits of the standard PETSc configuration, i.e., with 32 bit indices.

The current implementation of both discretization and model is tailored towards CPU based computing architectures.
However, we point out that most of this code has been prepared already with the perspective to be ported to GPU based computing architectures.
To achieve this goal there are however two main challenges.
First, the implementation of the quadrature rules which, due to the limited memory available and the GPU work model, requires specific design measures.
Second, the HJPGS algorithm has to be either ported to GPU (using independent-set coloring~\cite{zhang1996acceleration}), or a more suitable alternative with equivalent smoothing properties has to be found.
We emphasize that for remaining parts of our multilevel solver we can instead just switch to the back-end which targets GPUs, the Utopia/Tpetra backend.
Results presented in this work  are foreseen to be used for comparisons with future GPU accelerated versions of this code.

In this work we focus on networks with high fracture density, which represent a  challenging class of problems due to the complex geometry and the non-convexity of the underlying minimisation problem.
Future work shall include to port the entire framework to GPU architectures, and the integration of adaptive octree data structures  (e.g., by using DMPlex or P4est~\cite{BursteddeWilcoxGhattas11}) to efficiently handle the discrete representations of  sparse fracture networks.



\bibliographystyle{siamplain}
\bibliography{utopia}

\end{document}